\documentclass[10pt,twocolumn,letterpaper]{article}

\usepackage{cvpr}
\usepackage{times}
\usepackage{epsfig}
\usepackage{graphicx}
\usepackage{amsmath}
\usepackage{amssymb}
\usepackage{caption}
\usepackage{subcaption}
\usepackage{enumitem}

\newcommand\blfootnote[1]{%
  \begingroup
  \renewcommand\thefootnote{}\footnote{#1}%
  \addtocounter{footnote}{-1}%
  \endgroup
}

\usepackage[pagebackref=true,breaklinks=true,letterpaper=true,colorlinks,bookmarks=false]{hyperref}

\cvprfinalcopy 

\def\cvprPaperID{****} 

\ifcvprfinal\fi
\begin{document}

\title{Deep Learning-based Bio-Medical Image Segmentation using UNet Architecture and Transfer Learning}

\author{
Nima Hassanpour $^\dagger$\\
Georgia Institute of Technology \\
Department of Computer Science \\
{\tt nhassanpour3@gatech.edu}
\and
Abouzar Ghavami $^\dagger$\\
Georgia Institute of Technology \\
Department of Computer Science \\
{\tt apakdehi3@gatech.edu}
}

\maketitle

\blfootnote{The results and implementation of this paper was presented as a course project in the deep learning class at Georgia Institute of Technology, Atlanta, GA.}
\blfootnote{$^\dagger$ Authors have equal contribution to this paper.}


\begin{abstract}
   Image segmentation is a branch of computer vision that is widely used in real world applications including biomedical image processing. With recent advancement of deep learning, image segmentation has achieved at a very high level performance. Recently, UNet architecture is found as the core of novel deep learning segmentation methods. In this paper we implement UNet architecture from scratch with using basic blocks in Pytorch and evaluate its performance on multiple biomedical image datasets. We also use transfer learning to apply novel modified UNet segmentation packages on the biomedical image datasets. We fine tune the pre-trained transferred model with each specific dataset. We compare its performance with our fundamental UNet implementation. We show that transferred learning model has better performance in image segmentation than UNet model that is implemented from scratch.
\end{abstract}

\section{Introduction}
Convolutional neural network (CNN) had great impact on image processing and considerably improved accuracy of related tasks such as image classification, and object recognition \cite{al2018skin, nichols2019machine, shin2016deep} . But there are other type of tasks which need localization to highlight a specific feature or object in an image, which is called image segmentation. An image segmentation model assigns a label to all pixels of an image to locate objects inside an image. Image segmentation has application on wide range of fields such as transportation \cite{liu2018segmentation}, housing design \cite{wu2018boundary}, object detection \cite{garcia2018survey} and medical image \cite{al2018skin, chen2018drinet}, which we mainly focus on the last one.

An important and challenging task for biomedical researchers and radiologists is to recognize features in a medical images, like cancer cell or tooth cavity. Reading information from biomedical image requires to be a skillful specialist. In practice, it may not be even easy for an experienced specialist, therefore, having a model that highlights critical spots on medical image can be very helpful in diagnosis and treatment of diseases. A model that achieves this goal should have high accuracy, since it directly affects treatment procedure of patients. On the other hand, the proposed model should reach to high performance with relatively small datasets. Because annotating medical images needs skillful specialists and it is expensive to provide large number of samples. It is common to only have few hundreds of images as the training set of a segmentation model.

Several attempts has been made to tackle these challenges. Ciresan et al. \cite{ciresan2012deep} trained a network in a sliding-window setup to predict the class label of each pixel by providing a local region (patch) around that pixel as input. This model is localized and increases number of training dataset by using patches of images. But this strategy has two drawbacks. First the model is slow, and there is a trade off between localization and size of patches. Other works \cite{seyedhosseini2013image, hariharan2015hypercolumns} proposed classifier output that considers inputs from multiple layers, which they could improve Ciresan et al. approach to have good localization and use of context (patches) simultaneously.

But among the models that were proposed for image segmentation, UNet \cite{ronneberger2015u} had significant impact on this type of task, by generating a model that is fast, and localized while using the whole image as the input. Even after several years UNet is still building block of many new segmentation models \cite{siddique2021u} such as UNet++ \cite{zhou2018unet++}, attention UNet \cite{oktay2018attention} and so on. UNet uses fully convolutional model which includes contraction path (down-sampling), bottleneck connection (by CNN layers), and expansion path (up-sampling). In order to localization (i.e., assigning class label to each pixel), high resolution features from contraction path are combined with up-sampling outputs (residual connection). These innovations empower it to reach high accuracy even on fairly small training datasets. The UNet network is suitable for all kinds of biomedical segmentation problems and allows segmentation of arbitrarily large images. We discuss in detail about UNet architecture in section \ref{sec:approach}. Despite its significant improvements, UNet has two main limitation \cite{zhou2019unet++}. First its optimal depth (number of blocks for down-sampling) is unknown, and second, its skip connection imposes restriction to only aggregate the same-scale feature map of encoder and decoder sub-networks. More recent models such as UNet++ \cite{zhou2018unet++} address these problems with applying ensemble of UNets with different depth and designing skip connections with varying feature map for decoder.

\begin{figure}
\centering
\includegraphics[width=8cm]{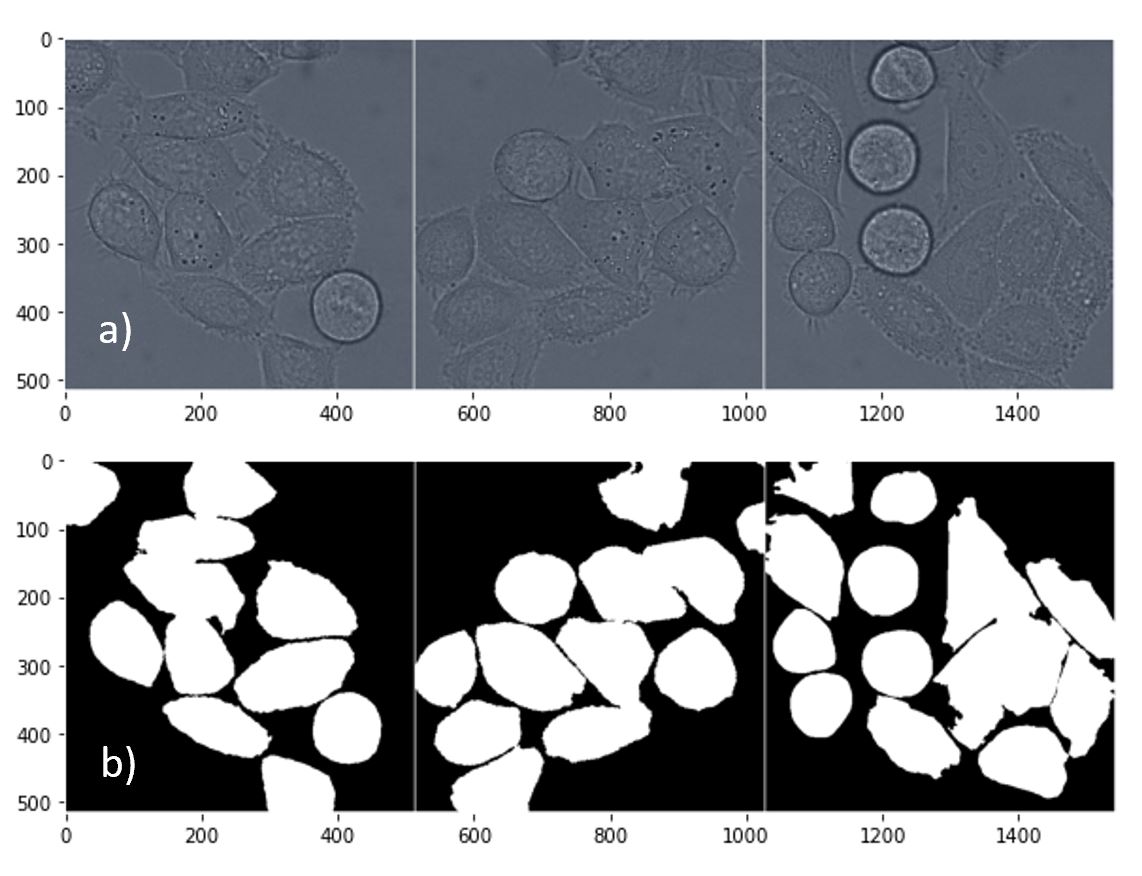} 
\caption{(a) Samples from HeLa cells. (b) binary masks for images in (a) that were segmented manually. } 
\label{pic:hela}
\end{figure}

\subsection{Dataset}
In this paper we use “DICHeLa” dataset that are images from HeLa cells on a flat glass recorded by differential interference contrast (DIC) microscopy. It contains 84 images and their mask as training dataset. Masks are manually annotated which shows the location of HeLa cells in the image. The datasets are found in the following link: \url{http://celltrackingchallenge.net/2d-datasets}. It is one of the datasets that were generated for cell tracking challenge which is part of International Symposium on Biomedical Imaging (ISBI) challenge.

Image \ref{pic:hela} shows samples of HeLa cells and their corresponding segmented masks. These masks are annotated manually, and the application of UNet is to generate these type of mask automatically for new images. As mentioned before, number of training dataset in biomedical imaging is low, and it this case, we are required to achieve to high quality segmentation with less than 100 training images.

\begin{figure}
\centering
\includegraphics[width=7.5cm]{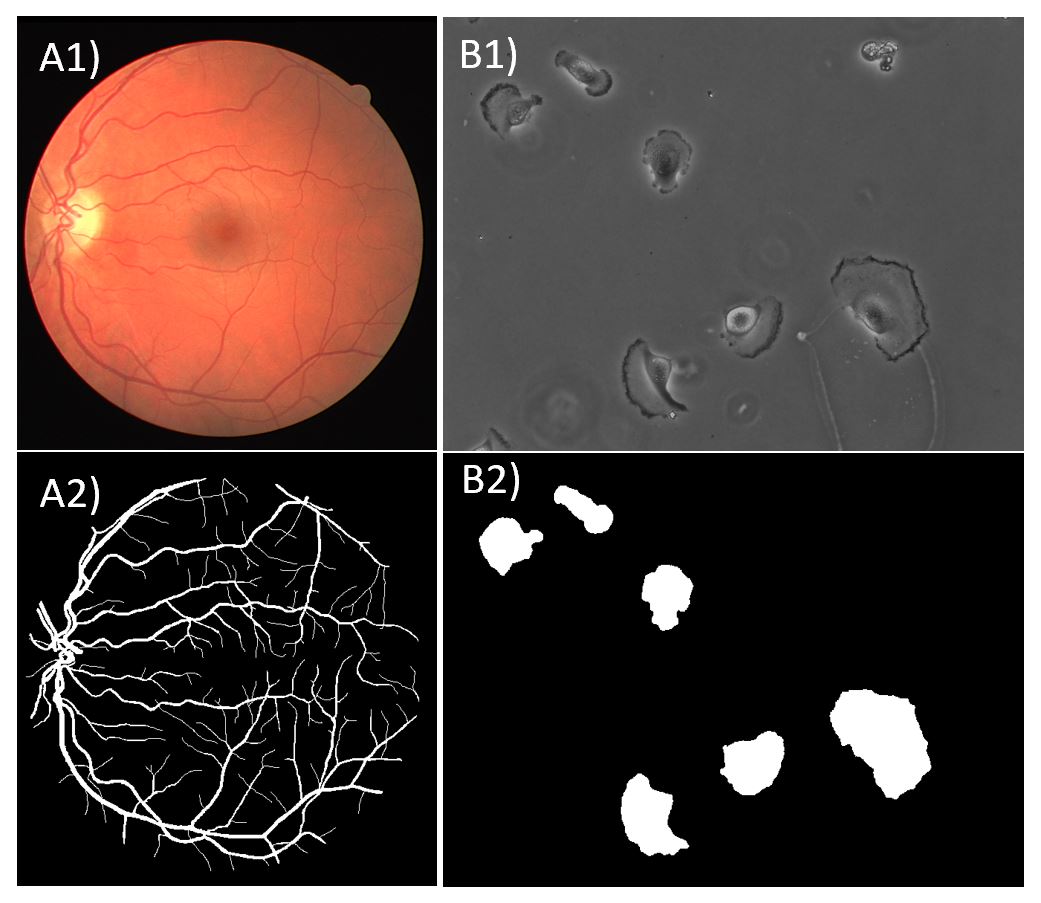} 
\caption{(A1-A2) Samples from retina and its vessel mask. (B1-B2) Image from PhC373 cell and its manually annotated segmented mask.} 
\label{pic:retina-phc}
\end{figure}

In order to proof application of UNet on different biomedical images we test it on two other datasets. PhC-C2DH-U373 dataset which is another dataset from cell tracking competition \cite{phc373_dataset}. It contains images from Glioblastoma-astrocytoma U373 cells on a polyacrylamide substrate, and they are relatively smaller than HeLa cells. This dataset includes 114 training samples and 114 images for testing. 

Digital Retinal Images for Vessel Extraction (DRIVE) is the other dataset we tested UNet on it \cite{retina_dataset}. Length, width, tortuosity, branching patterns of retinal blood vessel are utilized for the diagnosis, treatment, and evaluation of various cardiovascular diseases, diabetes, and hypertension. Automatic detection
of vasculature in retina helps on implementation of screening programs of those diseases. DRIVE dataset contains only 40 retina images and their vessel masks for both training and testing, which makes it a challenging segmentation task. The dataset can be found in the following link: \url{https://drive.grand-challenge.org/}. 
A sample from PhC373 and DRIVE datsets can be seen in Fig. \ref{pic:retina-phc}.

\section{Approach}\label{sec:approach}
In this paper we are looking for an advanced deep learning method to apply segmentation on image sets. We are looking for identifying objects in the scene. As a result we define two classes for each point of the output image. If the pixel contains the object, it is a class 1 pixel and if it does not contain the object and it belongs to background, it is in class 0. As a result, the input of our model is an image and the output is a mask image with two classes: $\{0, 1 \}$ for each pixel.

\begin{figure}
\centering
\includegraphics[width=0.5\textwidth]{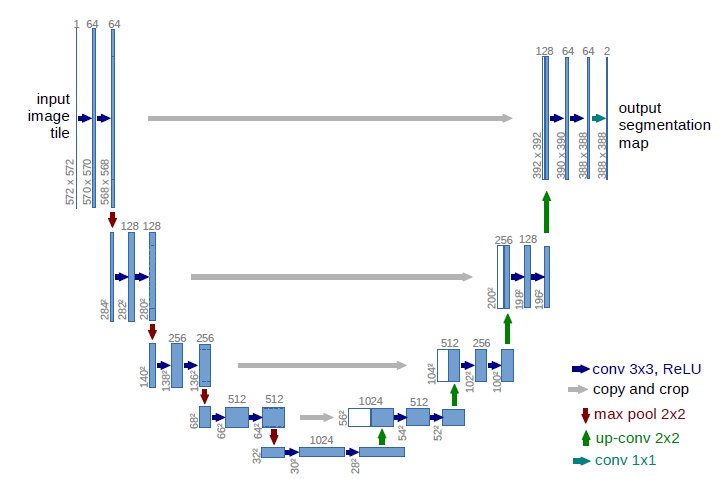}
\caption{UNet architecture - the core of state of the art deep learning-based segmentation methods (Courtesy of \cite{ronneberger2015u})}
\label{unet_architecture}
\end{figure}

We implement the UNET architecture that is proposed in \cite{ronneberger2015u}. UNET architecture contains two hierarchical (i) downsample and (ii) upsample modules (Fig \ref{unet_architecture}). Both dowansample and upsample modules have the same number of levels. The corresponding levels in each module are also connected through skip connections similarly used in ResNet \cite{resnet2016} architecture.

In downsample module, we apply two 2D-convolution layers with ReLU activation functions followed by a max-pooling layer. We also apply batch-normalization in each 2D convolution layer before ReLU activation. Before max-pooling, we also keep the output of each downsample module level to concatenate with the input of the same level image tensor in upsample module. 

In upsample module, we upsample the deconvolution layers using transpose convolution blocks. Each upsampled image tensor is concatenated with resized output of the same level from downsample module. The connection between output of each level of downsample module and input of the same level in upsample module is called skip connection. Note that adding skip connection helps UNet to find smaller size and detailed segments during training.

At our first try, we couldn't get acceptable results from UNet implementation. After many further investigation, we realized that the skip connection tensors are not transferred from upsample module to downsample module in backward training. It caused the incorrect and vague segmentation result. We used multiple Pytorch transformation functions including torchvision's centercrop and resize functions that guarantee the tensor is mutable during backward learning. The segmentation results were improved using mutable tensors. We also realized that resizing rather than cropping the output of downsample layers and pass them through skip connections leads into better segmentation results.

The authors in \cite{ronneberger2015u} use a specific morphological weighted softmax function to segment the specific cell datasets. This morphological transformation requires knowing the distance of each cell's pixel in the image to center of the cell and border of nearest adjacent cell. The paper does not reveal how these distances are calculated and just applies them as available data. In order to make the UNet structure more general for applying segmentation on any other datasets, we added a Sigmoid function in the last layer of upsample module for binary classification. It enabled us to apply our implemented UNet segmentation on other datasets such as Retina dataset for vessels' segmentation. 

Note that UNet structure is very dependent to the image size and it does not go deeper more than $\log \min(width, height)$ of image due to the max-pooling function with kernel size of $2$ that is used at the end of each level in the downsample module. In many situations, we need to evaluate our model with resized and downsampled version of original image to test its performance. So, we have written the architecture's code to be dynamically adaptive to the size of input image.

\section{Experiments and Results}\label{ExperimentSection}
\begin{figure}
    \centering
    \includegraphics[width=0.5\textwidth]{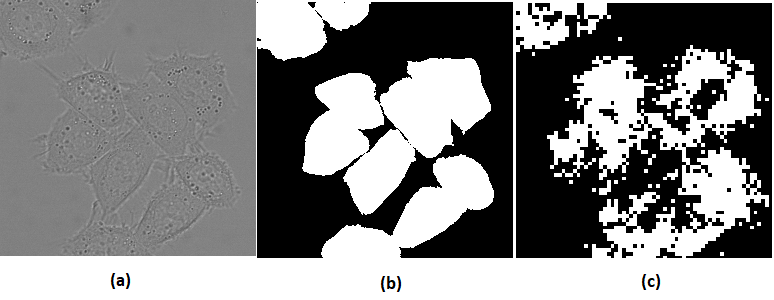}
    \caption{Result of our implemented UNet architecture on a sample from HeLa Cell dataset for cell segmentation: (a) Main image, (b) mask image and (c) predicted mask }
    \label{unet_scratch_HeLa}
\end{figure}

We applied segmentation tasks using our implemented UNet architecture on different biomedical datasets. There are two different sets of images with corresponding mask images in HeLa cell datasets each contains images of 84 HeLa cells. We considered the first set as the training set and the second set as validation set. Figure \ref{unet_scratch_HeLa} shows the results of our implemented UNet model on HeLa cell dataset for a test sample.
  We calculated a common Dice coefficient factor to match the results of our predicted segmentation and original mask image. The Dice coefficient is calculated as in the following:
  \begin{equation}
      \text{Dice Coeff.} = \frac{2.0 \times \text{Intersection of Binary Mask Images}}{\text{Union of Binary Mask Images}}
    \label{dice_factor_eqn}      
  \end{equation}
  
  Although our predicted segmentation result visually looks similar to the original mask picture, but we obtain a moderate average dice coefficient of 46 percent with respect to high average dice coefficient of 77 percent obtained in the original paper \cite{ronneberger2015u}. The reason is that the paper uses specific morphological transformation metric as described in Section \ref{sec:approach} that uses more internal information of provided data, while we use a more general cost function using Sigmoid at the last layer of the UNet architecture.
  
\begin{figure}
      \centering
      \includegraphics[width=0.5\textwidth]{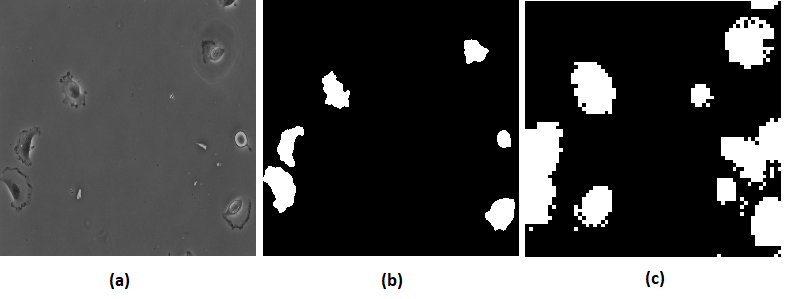}
      \caption{Result of our implemented UNet architecture on a sample from PhC373 cells' dataset for cell segmentation: (a) Main image, (b) binary mask image and (c) predicted mask.}
      \label{phc_sample_result}
  \end{figure}

  We also applied our implemented UNet structure on PhC373 cells dataset \cite{phc373_dataset}. Figure \ref{phc_sample_result} shows the result of our segmentation algorithm on a sample PhC373 image. As seen in Fig.  \ref{phc_sample_result}, the segmentation result does not visually predict the target mask for the sample PhC373 image. It shows that our general output UNet implementation does not ignore the small non-cell substances in the image and considers them as cell parts. The authors in \cite{ronneberger2015u} have used specific morphology transformation to consider only the cells. It seems that our general UNet model can not identify the small non-cell substances and we also need to remove them. We can remove these substances by using more specific data about PhC373 such as the cell-distance weighted cost function used in the last layer of \cite{ronneberger2015u}. The average Dice coefficient for PhC373 dataset is 41 percent that is much less than average Dice coefficient that is obtained in the paper \cite{ronneberger2015u} as 92 percent.
  
\begin{figure}
      \centering
      \includegraphics[width=0.5\textwidth]{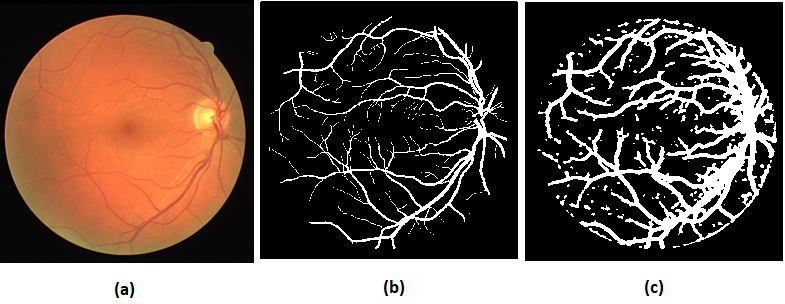}
      \caption{Result of our implemented UNet architecture on a sample from Retina dataset for vessel segmentation: (a) Main image, (b) binary mask image and (c) predicted mask.}
      \label{retina_sample_result}
  \end{figure}

  We also applied our generalized UNet implementation on segmentation of vessels in humans' Retina dataset \cite{retina_dataset}. Figure \ref{retina_sample_result} shows the result of our implemented UNet architecture on a sample of Retina dataset. As seen in Fig. \ref{retina_sample_result}, the result of our UNet application on Retina dataset for vessel segmentation is visually good. However, the average Dice coefficient for this image set is calculated as 34 percent that is low. The reason is that our UNet implementation results show the retina vessels very thicker than actual size. It causes the proportion of predicted binary image gets large in denominator of equation (\ref{dice_factor_eqn}) and makes the dice coefficient to be smaller than expected. 
  
  To improve the model, we used an Adam optimizer with learning rate set at $0.001$ . The learning rate adaptively changes by factor of $0.75$ at the mid and third-quarter of total number of epochs. In order to make the process to run faster, we used a resized version of input images and fed them to our implemented UNet model with sizes $256 \times 256$. We examined multiple loss functions and their mixtures including: Cross Entropy, Binary Cross Entropy, Focal Loss and Dice Loss. We obtained a relatively good results for our implemented UNet structure using mixed loss with the same weights of Dice coefficient loss and Focal loss with $gamma = 0.9$.
  
\begin{figure}
      \centering
      \includegraphics[width=0.5\textwidth]{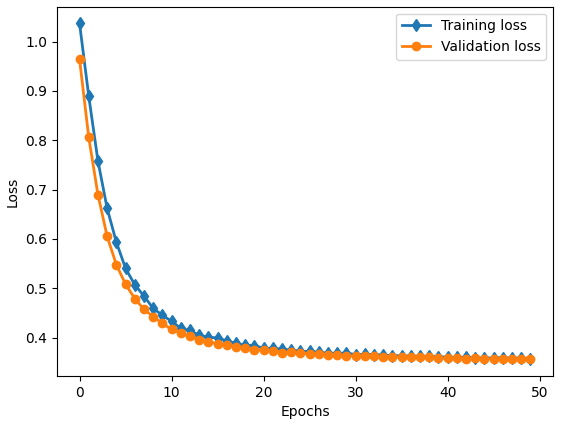}
      \caption{Training and validation loss of our implemented UNet on HeLa Cells dataset}
      \label{train_val_loss}
  \end{figure}

  Figure \ref{train_val_loss} demonstrates the training and validation loss of our model on HeLa cells dataset. It shows that training and validation loss is decreasing as the number of epochs increase. Furthermore, validation loss is less than training loss that shows our model didn't overfit. Our training loss stopped at $0.38$ that is not a very small value. It shows that our model accuracy is not very high that we also discussed its moderate Dice coefficient with respect to relatively high Dice coefficient that is obtained in the paper.
  
\subsection{Modifications for Result Improvements}
In this section we utilize several techniques to improve the performance of our model. Below are the main items were applied to increase accuracy of UNet:

\begin{itemize}[leftmargin=*]
   \item \textbf{Using pretrained models and initialize weights with transfer learning:} While exploring UNet performance, we found out initial weights plays an important role on accuracy of UNet. Accordingly, we decided to use initial weights from pretrained models that were trained on large image datasets. We implement it by using Segmentation models from segmentation models library \cite{smp_lib}. This library enabled us to use ResNet weights as the encoder, which has weights for training on ImageNet, semi-supervised learning (SSL) and semi-weakly supervised learning. 
    \item \textbf{Image augmentation:} Since the number of samples are low in segmentation tasks, image augmentation plays a key role in improving accuracy of the model. We applied horizontal flip and rotation to generate new samples. Other augmentation techniques were skipped to avoid overfitting.
    
    \item \textbf{Redefine loss function:} Loss function has a great impact on backpropagation procedure. A proper loss function adjust weights for predicted probability distribution of each label to increase probability of correct label. For this purpose, we combined focal loss and dice score as follows: $\alpha$(Focal Loss) - $\log$(Dice Score). Focal loss  is an improved version of Cross-Entropy loss (CE)  that tries to handle the class imbalance problem by assigning more weights to hard or easily misclassified examples and to down-weight easy examples (i.e. Background objects). Accordingly, with focal loss we increase importance of objects in image and reduce background pixels effect. On the other hand dice score calculates overlap pixels between predicted and ground truth masks. During training dice score should increase, and by subtracting dice score from focal loss having more overlap pixels means smaller loss function. Therefore, the loss function focus on segmentation of objects in image and increasing overlapping pixels simultaneously.
    
    \item \textbf{Hyper-parameter tuning:} After putting all previous states together, we optimzed the model by tuning hyper-parameters of the models. Parameters such as learning rate, $\gamma$ coefficient in focal loss, $\alpha$ parameter in mixed loss, batch size, number of epochs, image size, etc.. We tested different values for the hyper-parameters and re-trained the model with the set of hyper-parameters that gives the model highest dice score.
\end{itemize}

\begin{figure}
      \centering
      \includegraphics[width=0.5\textwidth]{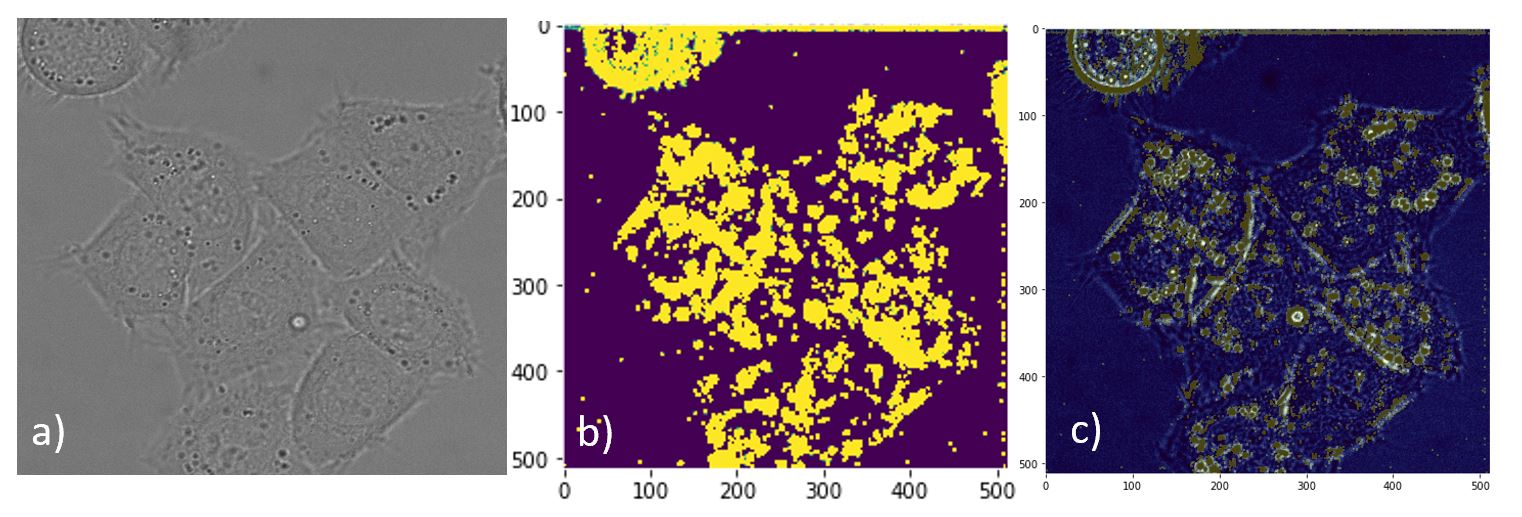}
      \caption{Improvement on HeLa dataset sample: (a) original image, (b) predicted mask image and (c)  predicted mask image merged with original image}
      \label{pic:imp_hela}
  \end{figure}

\begin{figure}
      \centering
      \includegraphics[width=0.5\textwidth]{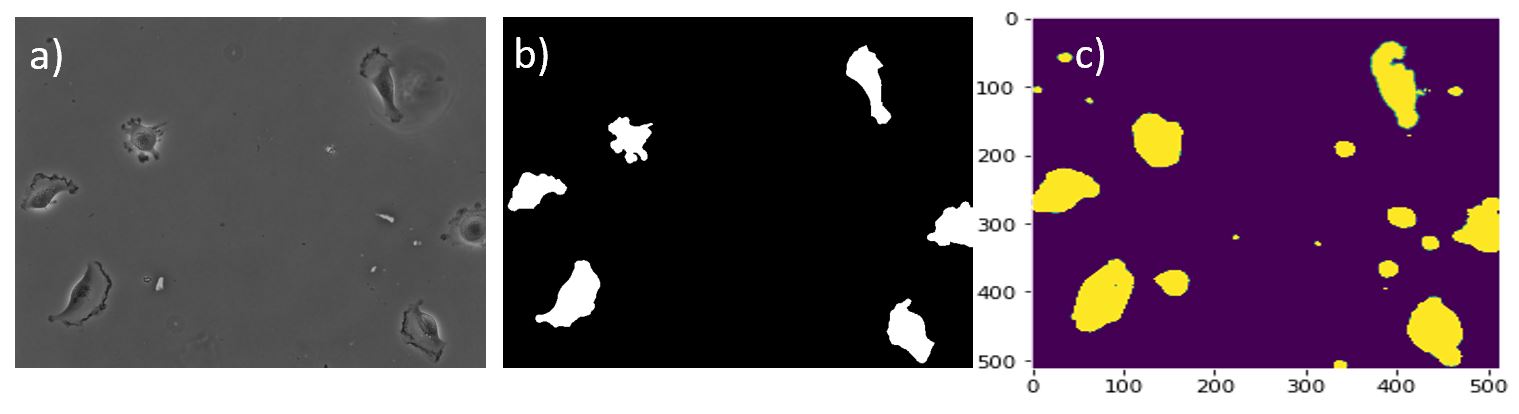}
      \caption{Improvement on PhC-U373 dataset sample: (a) original image, (b) binary mask image and (c) predicted mask image}
      \label{pic:imp_phc}
  \end{figure}

\begin{figure}
      \centering
      \includegraphics[width=0.5\textwidth]{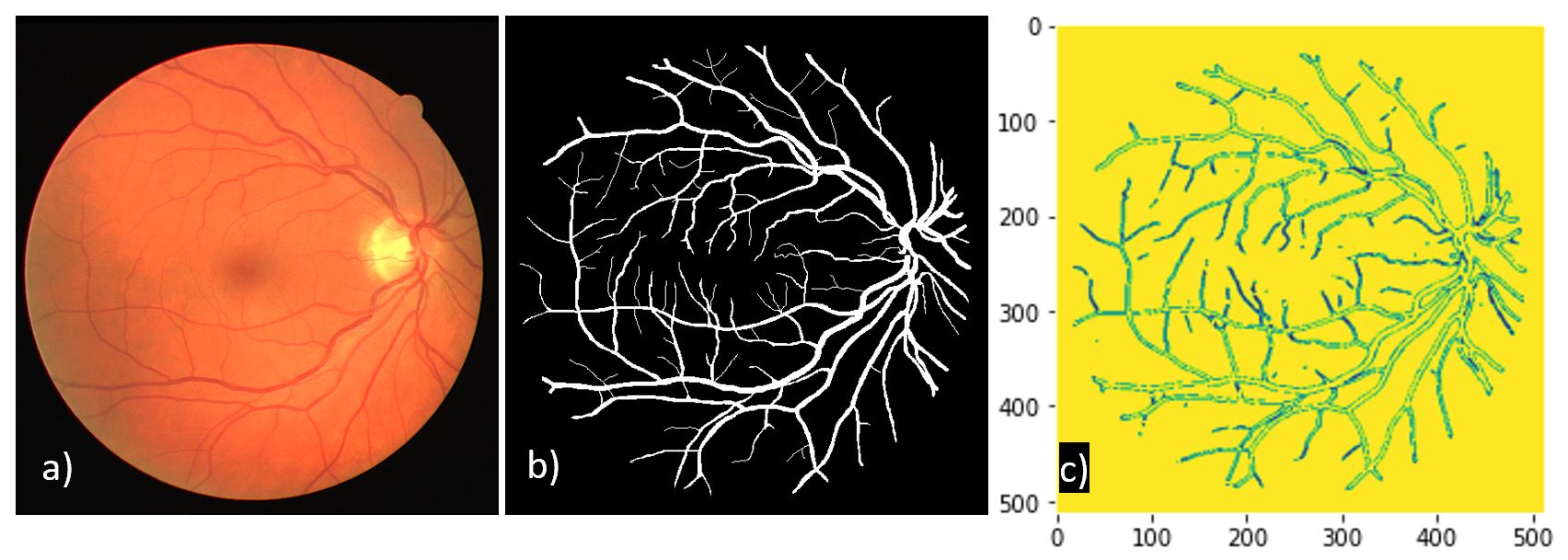}
      \caption{Improvement on Retina dataset sample: (a) original image, (b) binary mask image and (c) predicted mask image}
      \label{pic:imp_retina}
  \end{figure}

Figures \ref{pic:imp_hela}, \ref{pic:imp_phc}, \ref{pic:imp_retina} demonstrates the sample results from improved UNet model. As it is seen, all the predicted mask images visually match the corresponding target mask images. The improved predicted mask results are thinner than the predicted mask images obtained from our implemented UNet architecture. It shows that the improved pre-trained model qualitatively outperforms segmentation over from-scratch implemented UNet. 

Furthermore, the improved pre-trained model also quantitatively outperforms segmentation over our implemented UNet architecture. As table \ref{tab:comparison} demonstrates, Dice coefficients of improved pre-trained UNet model is much higher than the implemented UNet model for all three evaludated datasets. As discussed earlier, it shows that using Resnet encoder significantly outperforms traditional UNet performance. Also it shows weight initialization based on transfer learning from Imagenet dataset is critical in improvement of segmentation as we have access to limited number of images in our datasets.

\begin{table*}
\begin{center}
\begin{tabular}{|l|c|c|}
\hline
Dataset & Implemented UNet model & Improved pre-trained UNet model \\
\hline\hline
HeLa Cells & 46\% & 67\% \\
\hline
PhC-U373 & 41\% & 76\% \\
\hline
Retina & 34\% & 82\% \\
\hline
\end{tabular}
\end{center}
\caption{Dice coefficient comparison of from-scratch implemented UNet model and improved pre-trained UNet model}
\label{tab:comparison}

\end{table*}

\subsection{Conclusion}
We measure success both qualitatively and quantitatively. In qualitative term, our from-scratch implemented UNet architecture performs segmentation of the input image that is easily observed with human vision system. Our implemented UNet is not able to distinguish small substances inside image from target objects. For example, in PhC-U373 dataset, it segments non-cell small elements as well as cells. In Retina image, it shows vessels thicker than observed in binary mask image. In contrast, our improved pre-trained UNet model performs segmentation task in much more details.

From quantitative perspective, our from-scratch implemented UNet model shows small to moderate Dice coefficient. It is due to lack of detailed segmentation and rough object detection inside input image. However, our improved pre-trained model shows relatively high Dice coefficient along with high quality segmented images. 

As a result, we succeeded in our goal to perform deep learning-based segmentation with implementing UNet architecture from scratch and improving the pre-trained UNet model. We also applied these models on multiple datasets to make sure our models generalize segmentation in various real world scenarios.




\bibliographystyle{unsrt}


\end{document}